\documentclass[allclo]{FBSart}
\usepackage{amsfonts}
\usepackage{epsfig}
\usepackage{amssymb}
\newcommand{\be}{\begin{eqnarray}}
\newcommand{\ee}{\end{eqnarray}}
\title{Impact Parameter Dependent Parton Distributions for a Relativistic
Composite System }
\author{D. Chakrabarti\instnr{1,}\thanks{E-mail address: dipankar@phys.ufl.edu},
A. Mukherjee\instnr{2,}\thanks{E-mail address: 
asmita@lorentz.leidenuniv.nl}}
\instlist{Department of Physics, University of Florida, Gainesville, 
FL-32611-8440,USA \and Institute Lorentz,
University of Leiden, 2300 RA Leiden, Netherlands}

\runningauthor{D. Chakrabarti and A. Mukherjee}
\runningtitle{Impact parameter dependent parton distributions ....}
\begin{document}

\maketitle
\begin{abstract}
We investigate the impact parameter dependent parton distributions for a
relativistic composite system in light-front framework. We express
them in terms of overlaps of
light-cone wave functions for a self consistent 
two-body spin-$1/2$ state, namely  an electron dressed with a photon in QED.
The pdfs are distorted in the   
transverse space for transverse polarization of the state at one loop level.
\end{abstract}

\section{Introduction}
Impact parameter dependent parton distributions $q(x,b^\perp)$ 
\cite{bur1} have been
introduced recently as a physical interpretation of generalized parton
distributions (GPDs) in terms of probability densities in the
impact parameter space. When the state is transversely polarized, the 
impact parameter dependent pdf is distorted in the
transverse plane \cite{bur1}. 

Recently we have done an investigation of the pdf s in the impact parameter space for a relativistic 
composite system in the light-front
framework \cite{dip}, taking into account the correlation between different Fock
components of the light-cone (or light-front) wave function. We take
an effective spin ${1\over 2}$ system of an electron, dressed with a photon 
in QED. Such a model is self consistent and has been used
to investigate the helicity structure of a composite relativistic system
\cite{brod1}. The state can be expanded in Fock space in terms of light-cone
wave functions, which,  in this case can be obtained  
from perturbation theory, and thus their correlations are known at a certain
order in the coupling constant.  Earlier studies have shown that this gives
an intuitive picture of the DIS structure functions 
and scaling violations \cite{hari} and is suitable to address issues
related to the spin and orbital anglular momentum of the nucleon 
\cite{oam,hari1}. Such a state has also been used to investigate the 
twist three GPDs in terms of overlaps of light-cone wave functions \cite{mvh}.
We extend these studies to the impact parameter dependent
pdfs and here we report on our main results.

\section{Impact Parameter Dependent Parton Distributions}

For a transversely localized state \cite{bur1}
the impact parameter dependent pdfs are defined as
\be
q(x,b^\perp)=  \langle P^+, R^\perp=0^\perp, \lambda
\mid O_q(x,b^\perp) \mid P^+, R^\perp=0^\perp,\lambda \rangle
\label{impact}
\ee
with
\be 
O_q(x, b^\perp)= \int {dx^- \over 4 \pi} {\bar \psi} (-{x^-\over 2},
b^\perp)
\gamma^+ \psi({x^-\over 2}, b^\perp) e^{{i\over 2} x P^+ x^-}.
\ee
Instead of the fermion
operator, one can also have a gauge boson operator
\be
O_g(x, b^\perp)= \int {dx^- \over 4 \pi} F^{+ \nu}(-{x^-\over 2}, b^\perp)
{F^+}_\nu({x^-\over 2}, b^\perp) e^{{i\over 2} x P^+ x^-},
\ee
$b^\perp$ is the impact parameter, which is the transverse distance
of the active quark from the center of mass. We have taken the light-front
gauge, $A^+=0$. The transversely localized states are superpositions of
helicity eigenstates. It can be shown that $q(x, b^\perp)$ can be expressed as a
Fourier
transform of the GPD $H_q(x, 0, \Delta^2)$ \cite{bur1}:
\be
q(x,b^\perp)= {\mathcal H}_q (x,b^\perp) = \int {d^2 \Delta^\perp
\over (2 \pi)^2}  
e^{-i b^\perp. \Delta^\perp} H_q(x, 0, \Delta^2),
\ee
with
\be
H_q({x},0,\Delta^2)= \int {dz^-\over {8 \pi}} e^{{i\over 2}{x}{\bar
P} {z^-}}\langle P' \uparrow  \mid \bar \psi (-{ z^-\over 2})
\gamma^+\psi ({ z^-\over 2})
\mid P \uparrow \rangle.  
\label{def}
\ee
We take $\Delta^2$, the total momentum transfer to be purely
transverse. 

The helicity eigenstate of an electron dressed with a photon in QED can be 
expanded in Fock
space in terms of 1-body and 2-body light cone wave functions which can be 
expressed in terms of Jacobi momenta
$x_i, q_i^\perp$. The analytic form of the two-body wave function can be
obtained from the light-front eigenvalue equation \cite{hari}. The GPD
$H_{q,g}(x,0, \Delta^2)$ can be expressed as an overlap of the light-cone wave
functions. Fourier transform {\it w.r.t} $(\Delta^\perp)^2$ gives 
$q(x, b^\perp)$. The scale dependency in the impact parameter space comes
from the limits of the integration over the intrinsic transverse momenta of
the partons \cite{dip}. Integrating $H_q(x,,0)$ over $x$ one gets
$\int_0^1 dx H_q(x,0)= F_1(0)=1$, where
$F_1(0)$ is the form factor at zero momentum transfer. At 
non-zero $\Delta^\perp$, in the limit $x \rightarrow 1$, 
$H_{q,g}(x, \Delta^2)$ are
independent of $\Delta^\perp$. The impact parameter
dependent pdf in this limit is a delta function in $b^\perp$ as expected
because in this limit the electron carries all the momentum
and the transverse width of
the impact parameter dependent pdf vanishes \cite{bur1}.

\begin{figure}
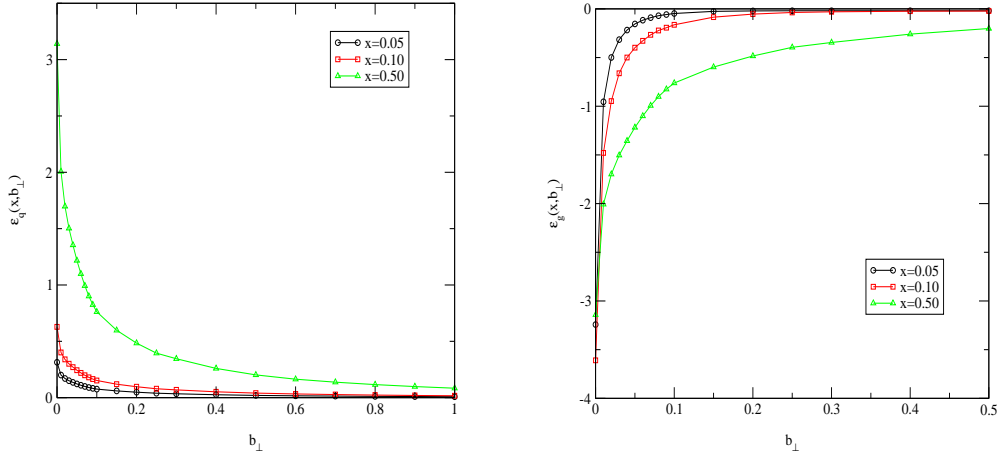

%\hspace{-0.3cm}
\parbox{7cm}{\epsfig{figure=Eq_x_.05_.1_.5.eps,width=6 cm,height=6 cm}}\
\
%\hspace{0.3cm}
\parbox{7cm}{\epsfig{figure=Eg_x_.05_.1_.5.eps,width=6 cm,height=6
cm}}\
\
\caption{${\mathcal E}_q (x, b^\perp)$ and 
$ {\mathcal E}_g (x, b^\perp)$ vs $b^\perp$ for three different  
values of $x$}
\end{figure}
\vspace{0.2cm}
 
The helicity flip part of the matrix element gives
\be
\int {dy^-\over 8 \pi} &&e^{{i\over 2} P^+ y^-x}\langle P+\Delta, \uparrow
\mid {\overline \psi}({-y^-\over 2}) \gamma^+ \psi ({y^-\over 2})\mid P,  
\downarrow \rangle =
{e^2\over (2 \pi)^3} \times \nonumber\\&&x (1-x)^2 (-i m) 
(-i \Delta^1-\Delta^2) \times   \nonumber\\&& \int
{d^2q^\perp \over {(q\perp^2+m^2 (1-x)^2) (q^\perp+(1-x) \Delta^\perp)^2
+m^2 (1-x)^2)}} \nonumber\\&&=-{E_q\over 2 m} (\Delta^1-i \Delta^2).
\ee
$m$ is the renormalized mass of the electron. There is a similar relation
for the gauge boson operator which gives $E_g(x,0,\Delta^2)$. 

Taking the Fourier transform, we get
\be
{\mathcal E}_{q,g}(x, b^\perp)&=&\int {d^2 \Delta^\perp\over (2 \pi)^2}
e^{-i b^\perp. \Delta^\perp} E_{q,g}(x,-(\Delta^\perp)^2). 
\label{fteq}
\ee
Fig. 1 shows the helicity flip contributions ${\mathcal E}_{q,g} 
(x, b^\perp)$ as a  
function of $b^\perp$ for three different values of $x$. The scale
dependence is suppressed in this case. We have plotted for 
positive $b^\perp$. ${\mathcal E}_q (x, b^\perp)$ is a smooth function of    
$b^\perp$ in the range shown and it increases as $b^\perp$ decreases. Also,  
it increases linearly with $x$. We have taken
the overall normalization ${\alpha\over 2 \pi}=1$ in order to study the
qualitative behavior and $m=0.5$. ${\mathcal E}_q (x, b^\perp)$ has a
maximum $b^\perp=0$. ${\mathcal E}_g(x, b^\perp)$ is negative for
positive $b^\perp$ and has a negative maximum at $b^\perp=0$. Like the
fermion case, ${\mathcal E}_g(x, b^\perp)$ is larger in magnitude for fixed
$b^\perp$ as $x$ increases. As before, we took ${\alpha\over 2 \pi}=1$ and 
$m=0.5$. When the state is transversely polarized, the derivative of  
${\mathcal E}_{q,g} (x, b^\perp)$ gives the distortion of the pdf in the
transverse space \cite{bur1}. The distortion of the distribution in impact
parameter space increases
as $b^\perp$ decreases and for a given $b^\perp$ the distortion is higher in
magnitude for larger values of $x$. The distortion shifts the distribution
${\mathcal E}_q (x, b^\perp)$ actually towards negative values of $b^\perp$.

\section{Summary}

We report on an investigation the impact parameter dependent parton 
distributions for a
relativistic composite system. An ideal framework is based on light-front
field theory, where the transverse boosts behave like  Galelian boosts and
the longitudinal boost operator produces just a scale transformation. We
take an effective composite spin $1/2$ state, namely an electron dressed
with a photon in QED. Using the overlap representation of GPDs in terms of
light-cone wave functions, we obtain the scale dependence of the impact 
parameter dependent pdfs at one loop. The helicity flip part gives the
distortion of the pdf in transverse space when the state is transversely
polarized.    

\begin{acknowledge}
We thank M. Burkardt for valuable discussions. 
AM thanks the organizers of Lightcone 2004 for a
wonderful and stimulating conference. The work of AM has been supported 
in part by the 'Bundesministerium f\"ur Bildung und Forschung', Berlin/Bonn.
\end{acknowledge}

\end{document}